\documentclass{article}

\usepackage{graphicx}
\usepackage{multirow}
\usepackage{textcomp}
\usepackage{siunitx}
\usepackage{amsfonts}
\usepackage{soul}
\usepackage{xcolor}
\usepackage{booktabs}
\sethlcolor{yellow}
\usepackage{graphicx, siunitx, subcaption}
\usepackage{authblk}

\usepackage[english]{babel}
\usepackage{textgreek}
\usepackage[backend=bibtex,style=numeric,sorting=none]{biblatex}

\usepackage[letterpaper,top=2cm,bottom=2cm,left=3cm,right=3cm,marginparwidth=1.75cm]{geometry}

\usepackage{amsmath}
\usepackage{graphicx}
\usepackage[colorlinks=true, allcolors=blue]{hyperref}
\title{{DSC} curve fingerprints directly encode mechanical properties of aluminum alloys}

\author[1]{Lukas Pichlmann}
\author[1,2]{Samuel Studer}
\author[1]{Aurel R. Arnoldt}
\author[1]{Paul Oberhauser}
\author[1]{Johannes A. Österreicher\thanks{Corresponding author: johannes.oesterreicher@ait.ac.at}}

\affil[1]{LKR Light Metals Technologies, AIT Austrian Institute of Technology, Ranshofen, Austria}
\affil[2]{Faculty of Science and Engineering, Maastricht University, Maastricht, The Netherlands}

\date{}
\begin{document}
\maketitle

\begin{abstract}
Differential scanning calorimetry (DSC) is a standard tool for studying precipitation and phase transformations in aluminum alloys, yet its relation to mechanical performance has so far remained mostly indirect. Here, we demonstrate that DSC curves themselves act as fingerprints that directly encode mechanical properties. Four representative 6xxx series alloys (Al--Mg--Si) were subjected to different natural and artificial aging regimens, followed by DSC heat-flow measurements and tensile testing. Machine learning models trained on the thermograms predicted yield strength, ultimate tensile strength, and uniform elongation in five-fold grouped cross-validation, with the best model (Lasso) achieving $R^2$ values of $0.93$, $0.86$, and $0.87$ and mean absolute errors of $14.3$~MPa, $11.1$~MPa, and $1.5$~\%, respectively. Leave-one-alloy-out evaluation with sparse calibration using anchor samples further demonstrated generalization across alloy chemistries. While direct prediction on unseen alloy data degraded performance substantially, inclusion of as few as one to two anchor conditions from the target alloy recovered predictive accuracy, approaching that of the standard cross-validation. Feature importance analysis revealed that the 230--270\textdegree C region, associated with precipitation of the primary hardening phase \textbeta$''$, contributed most strongly to predictive accuracy, providing direct mechanistic validation of the model. These findings establish DSC as a diagnostic tool that can serve as a rapid proxy for mechanical property evaluation, enabling accelerated alloy screening, process optimization, and integration of thermal analysis into data-driven manufacturing.
\end{abstract}

\section{Introduction}

Al–Mg–Si (6xxx series) alloys are among the most widely used aluminum alloys in transportation and construction, owing to their balanced combination of formability, corrosion resistance, and precipitation-hardened strength \cite{falkinger2024modeling}. Their mechanical performance is critically governed by the sequence and kinetics of precipitation reactions during natural and artificial aging, where the formation of metastable phases such as \textbeta$''$ provides the dominant strengthening contribution \cite{robson2020advances}. Understanding and controlling these precipitation processes is therefore essential for alloy development, process optimization, and quality assurance.

Differential scanning calorimetry (DSC) is a standard laboratory method for probing such precipitation and dissolution reactions in aluminum alloys \cite{kahlenberg2025deconvolution,Arnoldt2024,falkinger2022analysis,milkereit2015quench}. Exothermic peaks associated with the precipitation of aforementioned metastable strengthening phases, as well as endothermic dissolution signals, provide a thermal fingerprint of the evolving microstructure. These curves have been extensively used to identify transformation temperatures and precipitation sequences. However, their connection to mechanical properties has remained largely indirect: tensile strength and ductility are typically inferred only after complementary testing or by interpreting DSC features qualitatively. As a result, DSC has not yet been established as a stand-alone tool for predicting mechanical performance.

Prior studies have established correlations between DSC features and hardening behavior in Al--Mg--Si alloys. Natural and pre-aging treatments have been linked to changes in \textbeta$''$ precipitation and strength evolution \cite{Engler2019,Li2022}, while excess vacancies during artificial aging have been shown to influence DSC signatures and tensile properties \cite{Yang2021}. More recent work has exploited DSC peak analysis to interpret bake-hardening responses and natural-aging stability \cite{Li2022}, or combined DSC with complementary methods such as high-energy X-ray diffraction to resolve overlapping transformations \cite{Kahlenberg2024}. While these approaches confirm that DSC encodes valuable information about precipitation processes, they provide only indirect or assumption-driven links to mechanical performance.

Parallel advances in data-driven metallurgy demonstrate the feasibility of inferring mechanical properties directly from signal-type data, such as ultrasonic amplitude scans \cite{Park2023,park2025three} or X-ray diffraction profiles \cite{Murakami2025}. Physically-based models coupling DSC with tensile data have also been proposed \cite{Grohmann2026}, but these require explicit modeling of phase fractions. 

To our knowledge, no prior work has leveraged the full DSC thermogram as a direct input for machine learning to predict yield strength, ultimate tensile strength, and elongation in Al--Mg--Si alloys, a gap we address here. Variable importance analysis consistently highlights the region associated with \textbeta$''$ precipitation, providing direct insight into strengthening mechanisms without the need for exhaustive microstructural characterization. Furthermore, we demonstrate that models can be transferred to previously unseen compositions via domain adaptation with minimal calibration data.

\section{Materials and methods}

\subsection{Alloy compositions and heat treatments}

AA6016, AA6061, AA6063, and AA6082 were selected to cover a representative range of typical compositions within the 6xxx series of aluminum alloys, spanning variations in Si, Mg, Cu, and Mn contents, among other elements. The sheet thicknesses were \SI{1.0}{\milli\metre} for AA6061, \SI{1.2}{\milli\metre} for AA6063, and \SI{1.5}{\milli\metre} for AA6016 and AA6082. The chemical compositions, measured with a Spectromaxx LMX06 optical emission spectrometer, are presented in Table~\ref{tab:compositions}. All alloys meet standard requirements, with the exception of AA6063, where the Si content (0.64~wt.\%) slightly exceeds the specified maximum of 0.60~wt.\%.

\begin{table}[h!]
\centering
\caption{Chemical compositions of the alloys (wt.\%). ‘--’ indicates values below the limit of quantification (LOQ). 'Bal.' denotes the balance, i.e., the remainder of the alloy.}
\label{tab:compositions}
\begin{tabular}{lcccccccccc}
\toprule
Alloy & Si & Mg & Cu & Mn & Fe & Zn & Ti & Cr & Al \\
\midrule
AA6016 & 1.10 & 0.36 & 0.07 & 0.06 & 0.15 & -- & 0.02 & --  & bal. \\
AA6061 & 0.66 & 0.82 & 0.28 & 0.10 & 0.40 & 0.06 & 0.06 & 0.20  & bal. \\
AA6063 & 0.64 & 0.61 & 0.01 & 0.01 & 0.18 & 0.04 & 0.01 & --  & bal. \\
AA6082 & 0.89 & 0.64 & 0.05 & 0.46 & 0.46 & 0.03 & 0.02 & 0.01  & bal. \\
\bottomrule
\end{tabular}
\end{table}

Samples were subjected to solution heat treatment (SHT; \SI{530}{\celsius} or \SI{540}{\celsius} for at least \SI{20}{\minute}), quenched, and then exposed to various combinations of natural aging (NA) at room temperature  (RT) and artificial aging (AA) at elevated temperatures to generate a range of precipitation states and corresponding mechanical properties.

Natural aging after SHT was performed at room temperature for durations ranging from 1~hour to 23~days. Artificial aging treatments were conducted at temperatures between 180°C and 200°C for times ranging from 20~minutes to 6~hours either directly after SHT or after NA.

Additionally, the W temper (as-quenched) of AA6016 and AA6082, as well as the as-received (AR) T4 temper of AA6016, were investigated (see \cite{kaufman2013} for details on W and T4 temper designations). AA of AR AA6016-T4 was also performed (i.e., the only condition where we did not perform SHT), and some samples underwent a pre-aging (PA) treatment after SHT (prior to further AA).

Table~\ref{tab:treatments} summarizes the distribution of processing conditions across alloys and treatment categories, with $N_{\text{Cond}}$ and $N_{\text{DSC}}$ showing the number of conditions of a given category in the dataset and the number of DSC measurement specimens of a given category in the dataset, respectively.

The resulting dataset is somewhat heterogeneous, reflecting the fact that samples originate from multiple projects and theses with differing scientific objectives. From a machine learning perspective, such heterogeneity is not a drawback but rather beneficial, as it promotes learning across a broad range of processing paths and precipitation states. Crucially, all experiments were performed using an identical experimental methodology and data acquisition protocol, with machine learning–based analysis considered from the outset, ensuring comparability and minimizing confounding effects despite differences in sample provenance.

\begin{table}[ht]
\centering
\caption{Summary of aluminum alloy datasets, heat treatment conditions after SHT (except for AR samples, where no SHT was performed), and sample counts.}
\label{tab:treatments}
\begin{tabular}{llp{7cm}cc} 
\toprule
\textbf{Alloy} & \textbf{Category} & \textbf{Processing Details} & \textbf{$N_{\text{Cond}}$} & \textbf{$N_{\text{DSC}}$} \\
\midrule
\textbf{AA6061} & NA & \SI{22}{\day} & 1 & 3 \\
 & DAA & \SI{180}{\celsius} (\SI{1}{\hour}), \SI{200}{\celsius} (\SI{1}{\hour}; \SI{6}{\hour}) & 3 & 6 \\
 & NA+AA & \SI{22}{\day} NA+AA (AA conditions same as for DAA) & 3 & 6 \\
\midrule
\textbf{AA6063} & NA & \SI{22}{\day} & 1 & 3 \\
 & DAA & \SI{180}{\celsius} (\SI{1}{\hour}), \SI{200}{\celsius} (\SI{1}{\hour}; \SI{6}{\hour}) & 3 & 6 \\
 & NA+AA & \SI{22}{\day} NA+AA (AA conditions same as for DAA)  & 3 & 6 \\
\midrule
\textbf{AA6016} & W & as-quenched & 1 & 2 \\ 
 & NA & \SI{1}{\hour}; \SI{4}{\hour}; \SI{1}{\day}; \SI{8}{\day}; \SI{23}{\day} & 5 & 10 \\ 
 & DAA & \SI{180}{\celsius} (\SI{1}{\hour}); \SI{185}{\celsius} (\SI{35}{\minute}); \SI{200}{\celsius} (\SI{1}{\hour}; \SI{6}{\hour}) & 4 & 8 \\ 
 & NA+AA & \SI{8}{\day} NA+AA at \SI{180}{\celsius} (\SI{1}{\hour}); \SI{185}{\celsius} (\SI{35}{\minute}); \SI{200}{\celsius} (\SI{1}{\hour}; \SI{6}{\hour}) & 4 & 7 \\
 & AR; AR+AA & AR; AR+AA at \SI{180}{\celsius} (\SI{1}{\hour}); \SI{185}{\celsius} (\SI{35}{\minute}); \SI{200}{\celsius} (\SI{1}{\hour}; \SI{6}{\hour}) & 5 & 10 \\
 & PA; PA+AA & PA at \SI{100}{\celsius} (\SI{5}{\hour}); PA+AA at \SI{180}{\celsius} (\SI{1}{\hour}); \SI{185}{\celsius} (\SI{35}{\minute}); \SI{200}{\celsius} (\SI{1}{\hour}) & 4 & 8 \\
\midrule
\textbf{AA6082} & W & as-quenched & 1 & 2 \\
 & Direct AA & AA at \SI{180}{\celsius} (\SI{1}{\hour}; \SI{2}{\hour}; \SI{3}{\hour}; \SI{4}{\hour}; \SI{6}{\hour}); \SI{190}{\celsius} (\SI{1}{\hour}; \SI{3}{\hour}; \SI{6}{\hour}; \SI{9}{\hour}; \SI{12}{\hour}; \SI{24}{\hour}; \SI{48}{\hour}) & 12 & 19 \\
\midrule
\textbf{Total} & & & \textbf{50} & \textbf{96} \\
\bottomrule
\end{tabular}
\vspace{0.5em}\\
\footnotesize{NA: natural aging; AA: artificial aging; DAA: direct artificial aging; W: W temper; AR: as-received; PA: pre-aging.}

\end{table}

\clearpage
\subsection{DSC measurements}

DSC was carried out using a Netzsch DSC 204 F1 Phoenix device. All samples were milled from sheet material into disk-shaped specimens (\SI{4.5}{\milli\meter} in diameter) to avoid deformation effects associated with punching \cite{Arnoldt2024}. Measurements were performed between \SI{-50}{\degreeCelsius} and \SI{600}{\degreeCelsius} at a heating rate of \SI{10}{\kelvin\per\minute} and generally conducted in duplicate, with a small number of exceptions as noted in Table~\ref{tab:treatments}.
\subsection{Tensile testing}
Mechanical properties (0.2\% yield strength: YS, ultimate tensile strength: UTS, and uniform elongation: UE, all based on engineering stress–strain data) were obtained from tensile tests. Three tests were performed for each condition (four tests for AA6082), and the mean values were used as targets for the machine learning models. Flat tensile specimens were prepared according to DIN 50125-H12.5×50 and tested following ÖNORM-EN-ISO 6892-1B.

\subsection{Data preparation} 

\subsubsection{Preprocessing}

To remove instrumental artifacts and enable direct comparison across measurements from the recorded DSC thermograms, we implemented three main preprocessing steps: reference subtraction, baseline correction, and interpolation to a common temperature grid.

Reference subtraction and baseline correction was carried out in accordance with \textcite{Arnoldt2024}: First, a reference curve measured on high-purity aluminum (Al 99.98\%) was subtracted from each sample thermogram. This step removes contributions from the measurement system like crucible heat capacity and instrumental drift, isolating the heat flow signal associated with phase transformations in the alloy. Second, additional baseline correction was applied to account for residual curvature in the subtracted signal. Linear fits (and third-degree polynomial fits for 6016) were applied via least squares regression through two temperature regions where no significant precipitation or dissolution reactions are expected (one low- and one high-temperature region), and the resulting polynomial was subtracted from the entire curve. These regions were selected individually for each alloy system based on inspection of the thermograms, as peak positions in the curves vary with alloy composition and thermal history. Typical temperature regions for these fits were located at approximately 25--50°C (before the onset of accelerated clustering or Guinier--Preston zone formation) and 550--590°C (after major dissolution events). 


Finally, all corrected thermograms were interpolated onto a common temperature grid spanning 50°C to 580°C at 0.5°C intervals using linear interpolation, yielding 1061 data points per curve. Then, the discretized heat flow values could be used as input features for machine learning. Although the raw DSC curves spanned $-$50°C to 600°C, the temperature range was restricted to 50--580°C to exclude measurement artifacts, such as initial transient oscillations \cite{Arnoldt2024}, as well as the onset of melting in some curves.

\subsubsection{Feature engineering}
Prior to model training, we investigated several feature engineering and dimensionality reduction strategies to mitigate overfitting risk given the relatively small dataset size. Three manual approaches were evaluated: (i) fitting a fixed number of Gaussian functions to each thermogram and using the concatenated amplitudes, means, and standard deviations as feature vectors; (ii) a peak-finding algorithm that identified all peaks and troughs, computed the associated enthalpy via numerical integration, and used peak positions, onset and endset temperatures, and integrated enthalpies as features; and (iii) dividing the full temperature range into $n$ equal segments and using the integrated heat flow within each segment as input features. Out of these approaches, the segmented integration approach yielded the best predictive performance. However, the manual feature engineering methods in our tests were ultimately outperformed by applying PLS regression directly to the preprocessed thermograms, as described in the following Section. For readers interested in the details of our manual feature engineering efforts, we refer to our public Codeberg repository, formally archived on Zenodo \cite{pichlmann_2026_18936092}.


\subsubsection{Dimensionality reduction of raw thermograms} \label{dimred}

In addition to manual feature engineering, we evaluated principal component analysis (PCA) and partial least squares (PLS) as dimensionality reduction techniques applied to the raw, preprocessed DSC thermograms, resulting in a low-dimensional dataset that can directly be used for model fitting.

PCA is a widely used dimensionality reduction technique that projects high-dimensional input features $\mathbf{X} \in \mathbb{R}^{n \times p}$ (in our case, $p = 1061$ heat-flow values at different temperatures) onto a lower-dimensional space spanned by orthogonal components capturing maximal variance in the data \cite{jolliffe2016principal}. While PCA identifies directions of high variance, it does not take the target variables $\mathbf{Y}$ into account.

PLS projects the high-dimensional input features $\mathbf{X} \in \mathbb{R}^{n \times p}$ onto a lower-dimensional latent space while explicitly maximizing covariance with the target variables $\mathbf{Y} \in \mathbb{R}^{n \times q}$ (yield strength, ultimate tensile strength, and uniform elongation; $q = 3$). Specifically, PLS seeks weight vectors $\mathbf{w}_k$ and $\mathbf{c}_k$ such that the latent scores $\mathbf{t}_k = \mathbf{X}\mathbf{w}_k$ and $\mathbf{u}_k = \mathbf{Y}\mathbf{c}_k$ exhibit maximum covariance. The resulting latent variables (components) capture directions in feature space that are simultaneously high-variance and predictive of the targets \cite{wold2001pls}.

When using PLS, baseline correction of the raw thermograms improved prediction accuracy only marginally. This is expected, as PLS components are optimized to maximize covariance with the target properties and so systematic offsets or baseline drift that are uncorrelated with mechanical behavior contribute little to the latent representation and are effectively suppressed. Baseline correction was nevertheless retained, as it improves the physical interpretability of the thermograms and facilitates comparison across measurements.

\newpage~
\subsection{Model fitting and evaluation}

Following the dimensionality reduction, we evaluated multiple machine learning algorithms on the 20-dimensional latent PLS representation to predict all three mechanical properties (YS, UTS, and UE) simultaneously. Since we are working with a small data set, we excluded data hungry architectures and methods such as deep learning and limited the selection of models to classical machine learning models that tend to work better in low data regimes \cite{xu2023small}. The regressors were implemented using a multi-output framework using the Python package scikit-learn \cite{scikit-learn} and included Support Vector Regression (SVR), Random Forest (RF), Gradient Boosting (GB), as well as simple linear regression and its adaptations Ridge and Lasso regression.

\subsubsection{Hyperparameter selection} \label{paramsel}

Hyperparameter tuning was conducted via a grid search optimizing for the Mean Absolute Error (MAE). The search spaces encompassed regularization weights, margins, and kernel coefficients for SVR, ensemble sizes and tree architectures for RF and GB, and regularization penalties for Ridge and Lasso. Prior to the PLS transformation and regression steps, all input features were standardized to zero mean and unit variance within the respective cross-validation folds.

Within the grid search implementation, we employed a Repeated Group 5-Fold cross-validation strategy to ensure robust model evaluation and prevent data leakage between replicate measurements. The grouping variable was defined by the specific alloy and heat treatment condition, ensuring that identical processing states were kept strictly isolated in either the training or the validation fold. 

\subsubsection{Cross-validation protocol}

The final performance of the model and corresponding hyperparameters selected via grid search was again assessed using five-fold grouped cross-validation, which was repeated 100 times to quantify the variance arising from differently chosen folds. 
Within each fold, feature standardization (zero mean, unit variance) and PLS projection were strictly fitted only on the training data and subsequently applied to the held-out validation fold, to prevent any indirect data leakage.

\subsubsection{Generalization across different 6xxx alloys}

To assess whether models trained using our approach can also generalize across alloy chemistries rather than merely predicting within known compositions, we performed leave-one-alloy-out cross-validation. In each fold, all samples from one of the four alloy types (AA6016, AA6061, AA6063, AA6082) were held out as the validation set, while the remaining three alloys constituted the training set. The same preprocessing pipeline (standardization and PLS projection fitted on training data only) was applied.

This evaluation protocol represents a significantly harder task than the grouped cross-validation described before, as the model must extrapolate to an alloy with potentially different precipitation kinetics and peak positions. 

Additionally, the inclusion of 1--3 distinct processing conditions from the held-out alloy (anchor measurements) in the training set was evaluated. This was done to assess whether prediction accuracy could be improved through sparse calibration.

\subsubsection{Feature importance analysis} \label{vipsec}

As mentioned in section~\ref{dimred}, PLS seeks weight matrices $\mathbf{W}$ and $\mathbf{C}$ to create latent score matrices $\mathbf{T} = \mathbf{X}\mathbf{W}$ and $\mathbf{U} = \mathbf{Y}\mathbf{C}$ such that the covariance between $\mathbf{T}$ and $\mathbf{U}$ is maximized. Because our target matrix $\mathbf{Y}$ is multivariate, the algorithm fits a single set of latent components that jointly predict all three mechanical properties. This forces the model to learn the shared underlying physical dependencies between the thermal fingerprint and the overall mechanical performance, rather than fitting isolated models for each property.

To interpret the PLS model and identify which temperature regions most strongly dictate mechanical properties, we calculated variable importance in projection (VIP) scores. The VIP score quantifies the contribution of each input feature (temperature point) to the overall predictive model.

For the $j$-th feature, the VIP score is defined as:

$$\text{VIP}_j = \sqrt{p \sum_{k=1}^{K} \frac{SS_k (w_{jk}/\|\mathbf{w}_k\|)^2}{\sum_{k=1}^{K} SS_k}}$$

where $p$ is the total number of features ($p = 1061$ in our case), $K$ is the number of retained PLS components, $w_{jk}$ is the weight of feature $j$ in component $k$, and $SS_k$ is the sum of squared deviations of $\mathbf{Y}$ to its mean explained by component $k$.

The variables $\mathbf{w}_k$ and $w_{jk}$ define how the original temperature features map onto the latent space. The weight vector $\mathbf{w}_k$ represents the direction in the original feature space that maximizes covariance with the mechanical properties. The scalar $w_{jk}$ is the specific loading of the $j$-th temperature point within that vector. By normalizing $w_{jk}$ by the vector's norm $\|\mathbf{w}_k\|$, the squared term $(w_{jk}/\|\mathbf{w}_k\|)^2$ isolates the proportional contribution of the heat-flow measurement at that specific temperature to the $k$-th component. 

Because our PLS formulation targets the multivariate matrix $\mathbf{Y}$, the $SS_k$ term represents the variance explained across YS, UTS, and UE combined. Consequently, a high VIP score (typically considered to be $\text{VIP} > 1$) indicates a temperature region that contains critical information for the joint prediction of all three mechanical properties. This allows us to directly link predictive accuracy to the most mechanically impactful precipitation or dissolution events.

\section{Results and discussion}

\subsection{DSC results}

\begin{figure}[h!]
    \centering
    \includegraphics[width=0.8\textwidth]{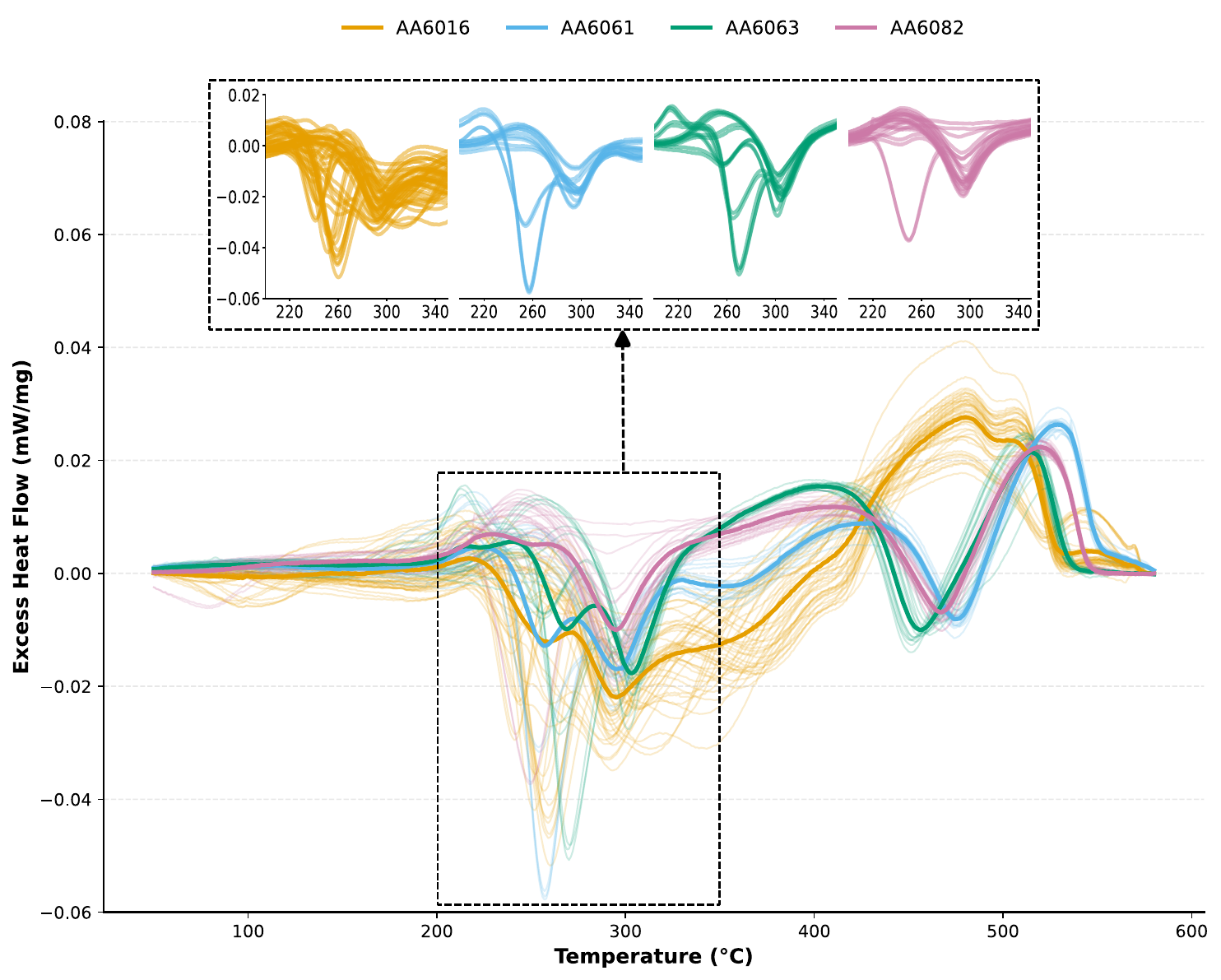}
    \caption{Baseline-corrected DSC thermograms for the four aluminum alloys in various tempers. Each narrow line represents an individual measurement, while bold lines show the mean curve of each alloy. For better readability, the magnified view shows the 200–350°C region for the individual measurements of each alloy.}
    \label{fig:curves}
\end{figure}

Figure~\ref{fig:curves} presents the DSC thermograms for all conditions investigated in this study. The characteristic exothermic and endothermic features associated with the precipitation and dissolution of metastable phases in Al–Mg–Si alloys are clearly visible for all four alloys~\cite{Arnoldt2024, kahlenberg2025deconvolution, Engler2019}. The temperature range between 200 and 350 °C roughly corresponds to precipitation reactions involving the principal strengthening phase \textbeta$''$ and subsequent phases such as \textbeta$'$ or B$'$ \cite{edwards1998}. For clarity, one curve per alloy and temper is also shown in the Appendix (Fig. \ref{fig:sparklines}).

\subsection{Feature selection}

\begin{figure}[h!]
\centering
\includegraphics[width=0.75\textwidth]{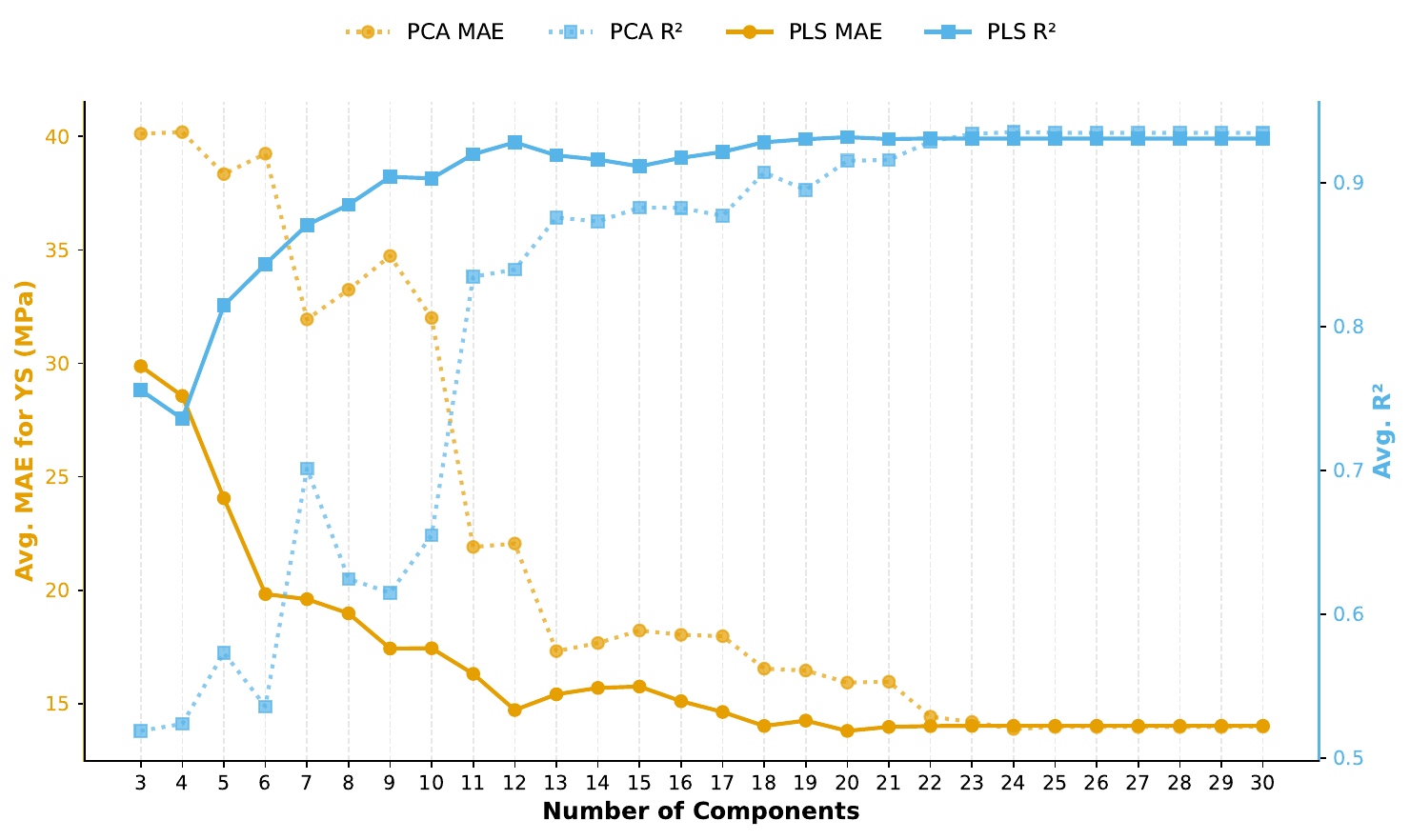}
\caption{Cross-validated prediction performance (on yield strength measurements) as a function of the number of latent components for PLS regression and PCA. Shown are the average $R^2$ (right axis) and mean absolute error (MAE) (left axis) scores across five folds.}
\label{fig:pls_vs_pca}
\end{figure}

Initially, we evaluated both principal component analysis (PCA) and partial least squares (PLS) as dimensionality reduction mechanisms for our model, and found that PLS consistently outperformed PCA across all target properties for limited number of components, though they converge as the number of components increases. This reflects the fact that PCA components are ranked solely by explained variance in $\mathbf{X}$, so predictive information may be distributed across many components rather than being concentrated in the leading components \cite{abdi2010partial}. Figure~\ref{fig:pls_vs_pca} displays model performance (of the final selected model) in cross validation for different numbers of latent components. Based on these cross-validation experiments, we selected $n_\mathrm{comp} = 20$ PLS components as the optimal number of latent dimensions, since performance at that point stagnates and no benefit is obtained by the inclusion of further PLS components into the feature set.

\subsection{Model selection and predictive performance}

\subsubsection{Five-fold grouped cross-validation}
We evaluated several machine learning algorithms to map the latent PLS representations to the target mechanical properties. Despite greater flexibility of nonlinear methods, none of these approaches significantly outperformed linear regression models on our dataset. 

The strong performance of linear models in this work does not imply that the underlying relationship between DSC heat flow and mechanical properties in 6xxx series alloys---or even in the four alloys considered here---is inherently linear. Rather, it may partly reflect the limited sample size ($n=96$), as small datasets typically favor simpler models, while more flexible learners are more prone to overfitting.

Table~\ref{tab:cv_results} summarizes the prediction performance of different regression models using PLS-reduced features in five-fold grouped cross-validation with the selected hyperparameters. For comparison, a naive baseline predictor is included in the table, which simply predicts the mean value of the corresponding target variables---yield strength (YS), ultimate tensile strength (UTS), and uniform elongation (UE)---in the training fold for all samples in the validation fold.

\begin{table}[h!]
\centering
\caption{Five-fold grouped cross-validation results using PLS dimensionality reduction ($n_\mathrm{components} = 20$, averaged over 100 repetitions). Shown are the means and standard deviations across all repetitions for each model. SVR stands for support vector regression.}
\label{tab:cv_results}
\resizebox{\textwidth}{!}{
\begin{tabular}{lcccccc}
\toprule
& \multicolumn{3}{c}{MAE} & \multicolumn{3}{c}{$R^2$} \\
\cmidrule(lr){2-4} \cmidrule(lr){5-7}
Model & YS (MPa) & UTS (MPa) & UE (\%) & YS & UTS & UE \\
\midrule
Baseline (Mean) & 69.2 $\pm$ 1.0 & 38.0 $\pm$ 0.7 & 5.6 $\pm$ 0.1 & -0.28 $\pm$ 0.28 & -0.24 $\pm$ 0.19 & -0.31 $\pm$ 0.43 \\
SVR & 17.6 $\pm$ 1.3 & 13.8 $\pm$ 1.0 & \textbf{1.4 $\pm$ 0.1} & 0.87 $\pm$ 0.03 & 0.79 $\pm$ 0.05 & \textbf{0.87 $\pm$ 0.04} \\
Random Forest & 27.4 $\pm$ 2.6 & 19.7 $\pm$ 1.5 & 2.1 $\pm$ 0.2 & 0.71 $\pm$ 0.10 & 0.60 $\pm$ 0.10 & 0.70 $\pm$ 0.12 \\
Gradient Boosting & 23.6 $\pm$ 2.3 & 17.5 $\pm$ 1.4 & 1.9 $\pm$ 0.2 & 0.77 $\pm$ 0.07 & 0.67 $\pm$ 0.09 & 0.74 $\pm$ 0.09 \\
Ridge & 15.1 $\pm$ 1.6 & 11.6 $\pm$ 1.3 & 1.8 $\pm$ 0.2 & 0.92 $\pm$ 0.03 & 0.85 $\pm$ 0.06 & 0.76 $\pm$ 0.10 \\
Lasso & \textbf{14.3 $\pm$ 1.4} & \textbf{11.1 $\pm$ 1.2} & 1.5 $\pm$ 0.1 & \textbf{0.93 $\pm$ 0.02} & \textbf{0.86 $\pm$ 0.06} & \textbf{0.87 $\pm$ 0.04} \\
Linear Regression & 15.2 $\pm$ 1.6 & 11.6 $\pm$ 1.3 & 1.8 $\pm$ 0.2 & 0.92 $\pm$ 0.03 & 0.85 $\pm$ 0.06 & 0.75 $\pm$ 0.10 \\
\bottomrule
\end{tabular}
}
\end{table}

Among the tested linear models, $L_1$-regularized regression (Lasso) achieved the best cross-validated performance, only slightly outperforming both unregularized ordinary least squares and $L_2$-regularized (Ridge) regression. The improvement over standard OLS-regression was modest, suggesting that the PLS projection already provides substantial regularization by restricting the model to a low-dimensional subspace.

The best value for the hyperparameter $\alpha$ of the Lasso model was determined as $\alpha = 1.0$ using the grid search approach outlined in section~\ref{paramsel}. In contrast to simple OLS regression, the added regularization parameter $\alpha$ controls the strength of the $L_1$ penalty applied to the regression coefficients. As a result, the model minimizes the objective function:

$$\min_{\beta} \left( \|\mathbf{y} - \mathbf{T}\beta\|_2^2 + \alpha \|\beta\|_1 \right)$$

with $\mathbf{T}$ representing the latent PLS features and $\beta$ the regression coefficients, solved independently for each target property. A high $\alpha$ differentiates the model from standard OLS by heavily penalizing large weights and forcing the coefficients of less predictive PLS components to zero. This induces strict sparsity, effectively performing a secondary feature selection on the latent variables and resulting in a simpler, more robust model \cite{tibshirani1996regression}.

\newpage
\subsubsection{Leave-one-alloy-out cross-validation}

To evaluate whether the model can generalize to entirely unseen alloy compositions, we performed leave-one-alloy-out (LOAO) cross-validation. In this more stringent protocol, all samples from one of the four alloys (AA6016, AA6061, AA6063, or AA6082) were held out as the validation set while the remaining three alloys formed the respective training set.

\begin{figure}[h!]
    \centering
    \includegraphics[width=\textwidth]{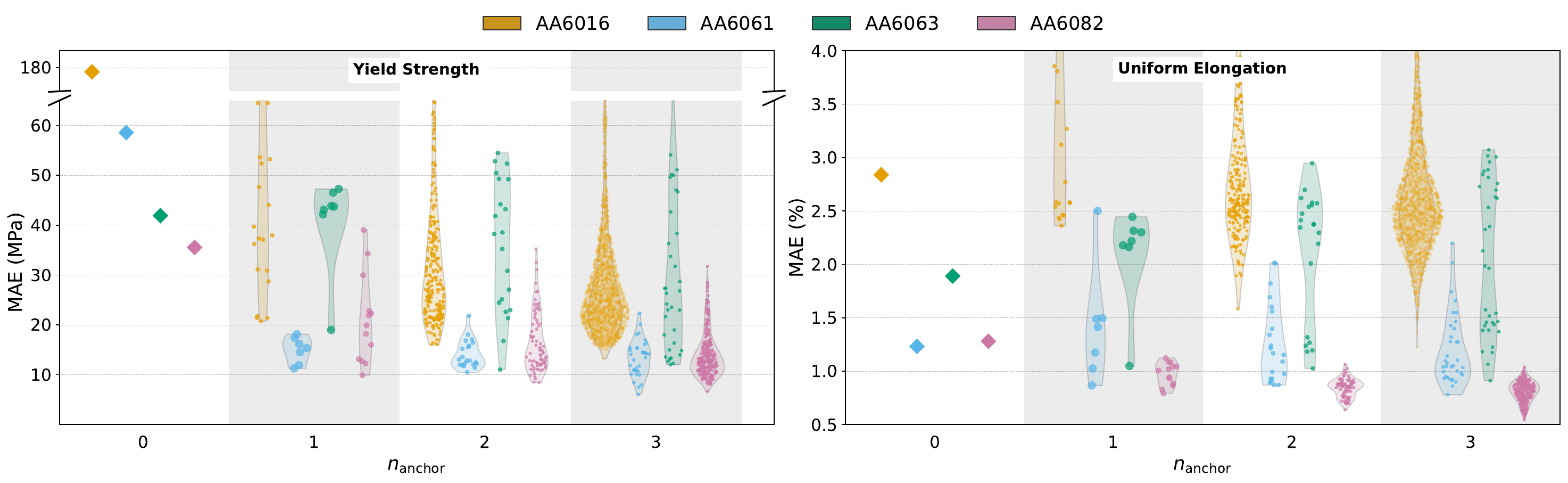}
    \caption{MAE-distribution for yield strength and uniform elongation as a function of the number of anchor conditions ($n_\mathrm{anchor}$) in leave-one-alloy-out cross validation. Scatter points show all possible combinations of anchor samples from the validation set.}
    \label{fig:loao_results}
\end{figure}

\paragraph{Direct transfer.} When no samples from the held-out alloy were included in training, model performance degrades substantially and becomes worse than even the naive baseline (Figure~\ref{fig:loao_results}, $n_\mathrm{anchor} = 0$). This indicates that DSC curve shapes do not transfer directly across alloy compositions and that each alloy exhibits distinct thermal signatures that require calibration.

\paragraph{Transfer with minimal target-domain data.} Performance recovered dramatically when even a small number of labeled samples from the target domain were included in training, consistent with supervised domain adaptation using minimal calibration data \cite{pan2009survey}. Each alloy composition defines a distinct domain characterized by its own precipitation kinetics and DSC signature distribution. The LOAO protocol thus constitutes a domain adaptation problem in which the model must generalize from source alloys to an unseen target alloy. With just one to two processing conditions from the held-out alloy added to the training set (and thus removed from the validation set), average mean absolute errors reduced drastically when using optimal anchor samples and started approaching or even improving upon the accuracy of the standard cross-validation.

\vspace{0.25cm}\noindent Figure~\ref{fig:loao_results} shows the mean average error (MAE) distribution for yield strength and uniform elongation as a function of the number of anchor conditions ($n_\mathrm{anchor}$) included in the training set, up to $n_\mathrm{anchor} = 3$. UTS resulted in a very similar distribution to YS, and was omitted for better readability (see Appendix, Figure~\ref{fig:loao_results_uts} for the distribution of UTS errors). For $n_\mathrm{anchor} > 0$, the predictive performance depends on the specific processing conditions selected as anchors. To capture this variance, an exhaustive evaluation of all possible combinations of anchor samples was performed, visualized as scatter points along with the violin plots. The steepest improvement occurs between zero and one anchor condition, after which performance gains diminish. This pattern was consistent across all four alloys, although the absolute performance varied depending on how similar the held-out alloy's DSC signatures were to those in the training set. The large errors observed for AA6016, especially for $n_\mathrm{anchor} = 0$, is due to this alloy type making up almost half of the samples in the dataset (see Table~\ref{tab:treatments}), resulting in a significantly reduced training set when using AA6016 as validation fold and thus insufficient data to reliably fit the model parameters. Further, AA6063 showed notably high sensitivity to the specific anchor samples selected for calibration, displaying a large variance in MAE with only a small subset of anchor samples leading to markedly improved performance.

Metallurgically, this comparatively low performance for AA6063 can be attributed to its negligible Cu content combined with lower Mg and Si levels (Table~\ref{tab:compositions}). These compositional differences lead to slower kinetics or different precipitation pathways compared to the other, higher-alloyed alloys. In particular, the negative impact of natural aging on subsequent artificial aging diminishes in lower-alloyed Al–Mg–Si alloys and can even become slightly positive in very lean compositions \cite{banhart2010natural,lai2017low,madanat2025heat}. We suggest that performance may improve when at least one compositionally similar alloy is present in the training dataset. Alternatively, it may be possible to actively select optimal anchor conditions, and further research into the characteristics that define effective anchors is warranted.

The results show that a sparse calibration set is often sufficient to adapt the model to a new 6xxx alloy composition. This has practical implications for alloy screening: rather than requiring exhaustive mechanical testing of every processing variant, a minimal set of anchor measurements enables high-throughput DSC-based prediction across the remaining conditions.

\newpage
\subsection{Feature importance analysis}

\begin{figure}[h!]
    \centering
    \includegraphics[width=\textwidth]{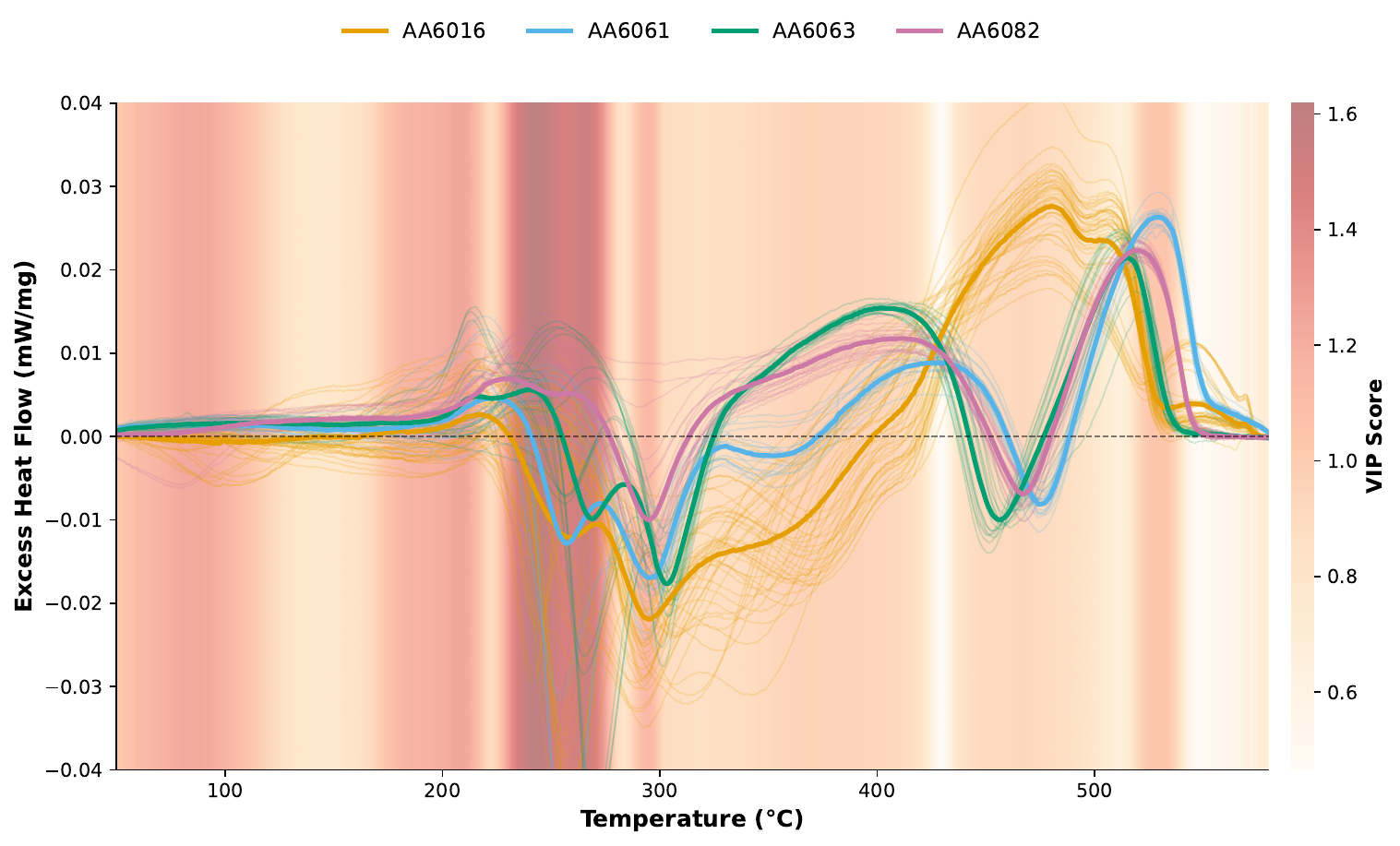}
    \caption{Variable Importance in Projection (VIP) scores as a function of temperature depicted as shaded background. The background intensity highlights the most predictive region. In general, VIP scores exceeding a threshold of 1.0 are deemed significant. The most impactful region is shown to occur between 230--270°C.}
    \label{fig:vip_scores}
\end{figure}

To understand which measurement values drive the predictive capabilities of the PLS-based regression model, we analyzed the variable importance in projection (VIP) scores, as introduced in section \ref{vipsec}.

Figure~\ref{fig:vip_scores} displays the VIP scores as a function of temperature. The dominant feature importance region spans approximately 230--270°C, where VIP scores reach their maximum. This temperature range corresponds to the  precipitation and/or dissolution of the \textbeta$''$ phase, which is widely recognized as the primary strengthening precipitate in Al-Mg-Si alloys \cite{marioara2024atomic}. The prominence of this region provides direct mechanistic validation of the model, showing that the thermal signatures most predictive of mechanical performance are precisely those associated with the thermal signature of coherent, high-number-density \textbeta$''$ precipitates that constitute the dominant obstacle to dislocation motion in peak-aged 6xxx alloys.

\begin{figure}[h!]
    \centering
    \includegraphics[width=\textwidth]{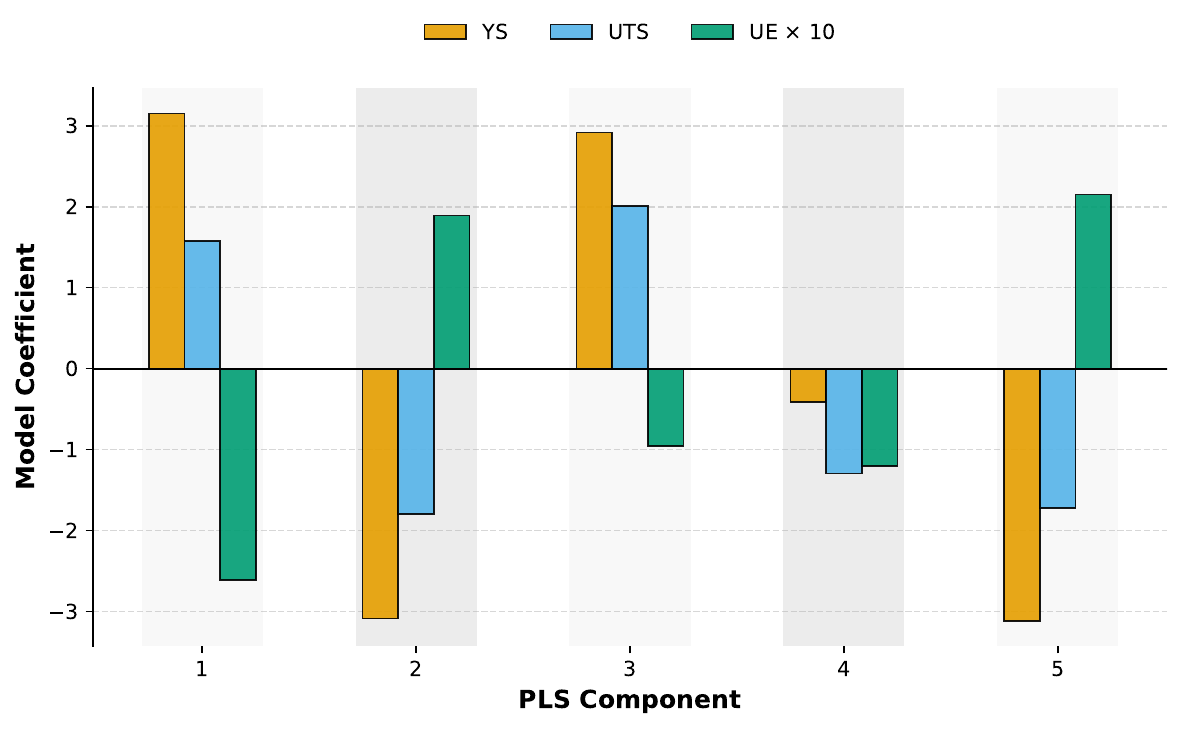}
    \caption{Lasso regression coefficients for the top 5 PLS components across the three mechanical properties. UE coefficient values are multiplied by 10 for better readability.}
    \label{fig:lasso_coef}
\end{figure}

Figure~\ref{fig:lasso_coef} displays the Lasso regression coefficients for the top 5 PLS components of the Lasso model fitted on the whole dataset. Yield strength and ultimate tensile strength exhibit similar coefficient patterns across components, consistent with the strong positive correlation between these two properties. In contrast, uniform elongation shows opposite signs on most components, reflecting the strength-ductility tradeoff in precipitation-hardened aluminum alloys \cite{poole2005}.

\newpage~

\section{Conclusion}

This work explores a novel approach to aluminum alloy characterization by demonstrating that the entire differential scanning calorimetry (DSC) thermogram can be utilized as direct modeling input to quantitatively predict mechanical performance, establishing DSC as a predictive tool rather than just a qualitative microstructural probe.

Specifically, we show that DSC thermograms directly encode microstructural information governing the mechanical properties of Al–Mg–Si alloys. Simple regression models can predict yield strength, ultimate tensile strength, and elongation from the thermograms alone, with feature importance analysis showing that \textbeta$''$ precipitation signatures dominate predictions, consistent with their central strengthening role.

While direct cross-alloy transfer without target-domain data fails, calibration with just one to two anchor conditions enables accurate domain adaptation, highlighting a common low-dimensional structure in DSC fingerprints across 6xxx alloys. This approach provides an alternative to conventional tensile testing, supporting accelerated batch screening \cite{vecchio2021}, heat-treatment optimization, and property assessment in geometrically constrained locations.

Despite these promising results, the current framework has boundaries that must still be addressed before industrial deployment. Current limitations include the limited dataset with a restricted set of alloy compositions, aging conditions, and a single DSC instrument and heating rate; future work may extend the methodology to other alloy series and explore active anchor selection, as well as the use of more complex non-linear models on extended datasets. Additionally, coupling these data-driven models with physically based approaches could enable the inverse design of aging schedules and broaden overall predictive capabilities.

\section*{Data Availability}

All data and code used to generate the results presented in this article have been deposited in a public repository on Codeberg and Zenodo, available at \url{https://codeberg.org/LKR/DSCorr} and \url{https://doi.org/10.5281/zenodo.18936092}. The repository includes the (processed) DSC curves, mechanical property data, feature extraction scripts, and machine learning model code necessary to reproduce the analyses and figures in this manuscript.

\section*{Author Contributions}

L.P. conceived and developed the machine learning methodology, implemented the software, performed the formal analysis, curated the data, and wrote the original draft. S.S. contributed to the experimental investigation, data curation, and methodology development. P.O. carried out experimental investigation and data curation. A.R.A. carried out experimental investigation. J.A.Ö. conceptualized the study, contributed to the methodology, provided resources, acquired funding, supervised the work, and contributed to writing the original draft.


\section*{Acknowledgments}

This work was supported by the ProMetHeus project, funded under COMET by the Austrian Ministry of Climate Action, Environment, Energy, Mobility, Innovation, and Technology (BMK), the Ministry of Labour and Economy (BMAW), and the funding agencies of Upper Austria, Styria, and Lower Austria—conducted by the Austrian Research Promotion Agency (FFG) [grant No. 904919], and co-funded by the European Union under the Horizon Europe COMPASS project [grant number 101136940]. Views and opinions expressed are those of the authors only and do not necessarily reflect those of the European Union, the European Commission, or the granting authorities. Neither the European Union, the European Commission, nor the granting authorities can be held responsible for them.

\appendix

\renewcommand{\thetable}{A\arabic{table}}
\setcounter{table}{0} 

\setcounter{figure}{0}
\renewcommand{\thefigure}{A\arabic{figure}}

\newpage

\section*{Appendix}
\subsection*{Individual DSC measurements}
\begin{figure}[h!]
    \centering
    \includegraphics[width=\textwidth]{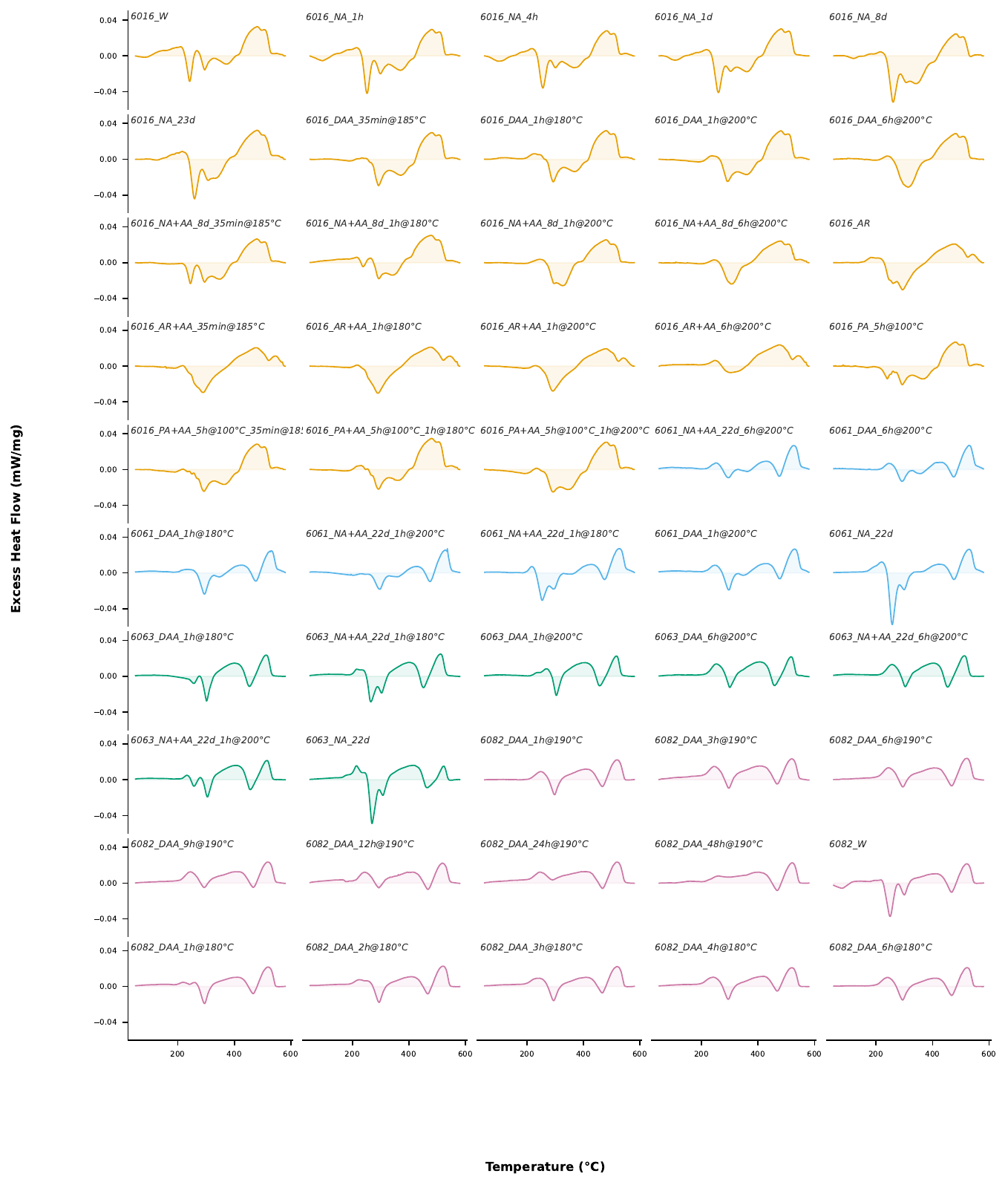}
    \caption{DSC curves for each unique condition in our dataset, with replicate conditions omitted.}
    \label{fig:sparklines}
\end{figure}

Figure~\ref{fig:sparklines} shows all baseline-corrected DSC thermograms of our dataset individually, for better readablity. Replicate DSC measurements are not displayed in this Figure.

\subsection*{Raw MAE values in leave-one-alloy-out cross-validation}

Table~\ref{tab:loao_results_per_alloy} shows the average performance in LOAO cross-validation for all possible combinations of anchor samples per held out alloy. 
\begin{table}[h!]
\centering
\caption{Leave-one-alloy-out cross-validation results for Lasso regression. Results are grouped by test alloy. $n_\mathrm{anchor}$ denotes the number of distinct processing conditions from the held-out alloy included in training. Values are Mean $\pm$ standard deviation across all possible combinations of anchor conditions for the given value of $n_\mathrm{anchor}$ (extended to $n_\mathrm{anchor} = 4$ beyond the $n_\mathrm{anchor} \leq 3$ shown in the main figures for completeness).}
\label{tab:loao_results_per_alloy}
\begin{tabular}{lcccccc}
\toprule
& & & & \multicolumn{3}{c}{Avg. MAE} \\
\cmidrule(lr){5-7}
Alloy & $n_\mathrm{anchor}$ & $N_\mathrm{train}$ & $N_\mathrm{test}$ & YS (MPa) & UTS (MPa) & UE (\%) \\
\midrule
\multirow{4}{*}{6016} & 0 & 51 & 45 & 178.25 & 96.73 & 2.84 \\
& 1 & 53.0 $\pm$ 0.2 & 43.0 $\pm$ 0.2 & 45.78 $\pm$ 19.86 & 28.14 $\pm$ 9.52 & 3.65 $\pm$ 1.43 \\
& 2 & 54.9 $\pm$ 0.3 & 41.1 $\pm$ 0.3 & 32.91 $\pm$ 13.61 & 24.53 $\pm$ 7.18 & 3.06 $\pm$ 0.98 \\
& 3 & 56.9 $\pm$ 0.3 & 39.1 $\pm$ 0.3 & 28.00 $\pm$ 10.33 & 22.37 $\pm$ 6.30 & 2.73 $\pm$ 0.71 \\
& 4 & 58.8 $\pm$ 0.4 & 37.2 $\pm$ 0.4 & \textbf{25.18 $\pm$ 8.36} & \textbf{20.82 $\pm$ 5.60} & \textbf{2.53 $\pm$ 0.56} \\
\cmidrule{1-7}
\multirow{4}{*}{6061} & 0 & 81 & 15 & 58.55 & 28.14 & 1.23 \\
& 1 & 83.1 $\pm$ 0.4 & 12.9 $\pm$ 0.4 & 14.96 $\pm$ 2.61 & 9.62 $\pm$ 4.69 & 1.42 $\pm$ 0.53 \\
& 2 & 85.3 $\pm$ 0.5 & 10.7 $\pm$ 0.5 & 14.07 $\pm$ 2.80 & 8.33 $\pm$ 2.37 & 1.27 $\pm$ 0.38 \\
& 3 & 87.4 $\pm$ 0.5 & 8.6 $\pm$ 0.5 & 13.29 $\pm$ 3.59 & 8.19 $\pm$ 1.83 & 1.20 $\pm$ 0.33 \\
& 4 & 89.6 $\pm$ 0.5 & 6.4 $\pm$ 0.5 & \textbf{12.70 $\pm$ 4.71} & \textbf{7.92 $\pm$ 1.96} & \textbf{1.15 $\pm$ 0.30} \\
\cmidrule{1-7}
\multirow{4}{*}{6063} & 0 & 81 & 15 & 41.91 & 37.59 & \textbf{1.89} \\
& 1 & 83.1 $\pm$ 0.4 & 12.9 $\pm$ 0.4 & 40.79 $\pm$ 9.79 & 33.90 $\pm$ 8.55 & 2.10 $\pm$ 0.47 \\
& 2 & 85.3 $\pm$ 0.5 & 10.7 $\pm$ 0.5 & 35.80 $\pm$ 13.28 & 31.36 $\pm$ 11.67 & 2.11 $\pm$ 0.62 \\
& 3 & 87.4 $\pm$ 0.5 & 8.6 $\pm$ 0.5 & 29.22 $\pm$ 14.89 & 26.18 $\pm$ 13.80 & 2.05 $\pm$ 0.71 \\
& 4 & 89.6 $\pm$ 0.5 & 6.4 $\pm$ 0.5 & \textbf{23.78 $\pm$ 13.99} & \textbf{20.51 $\pm$ 13.89} & 1.94 $\pm$ 0.80 \\
\cmidrule{1-7}
\multirow{4}{*}{6082} & 0 & 75 & 21 & 35.52 & 29.64 & 1.28 \\
& 1 & 76.6 $\pm$ 0.5 & 19.4 $\pm$ 0.5 & 20.95 $\pm$ 8.90 & 15.32 $\pm$ 6.68 & 0.97 $\pm$ 0.11 \\
& 2 & 78.2 $\pm$ 0.7 & 17.8 $\pm$ 0.7 & 16.55 $\pm$ 5.95 & 11.94 $\pm$ 4.59 & 0.84 $\pm$ 0.09 \\
& 3 & 79.8 $\pm$ 0.8 & 16.2 $\pm$ 0.8 & 13.92 $\pm$ 4.44 & 10.60 $\pm$ 3.36 & 0.81 $\pm$ 0.09 \\
& 4 & 81.5 $\pm$ 0.8 & 14.5 $\pm$ 0.8 & \textbf{12.30 $\pm$ 3.58} & \textbf{9.89 $\pm$ 2.44} & \textbf{0.80 $\pm$ 0.12} \\
\bottomrule
\end{tabular}
\end{table}

\subsection*{Metrics of leave-one-alloy-out cross validation for UTS}
\begin{figure}[h!]
    \centering
    \includegraphics[width=\textwidth]{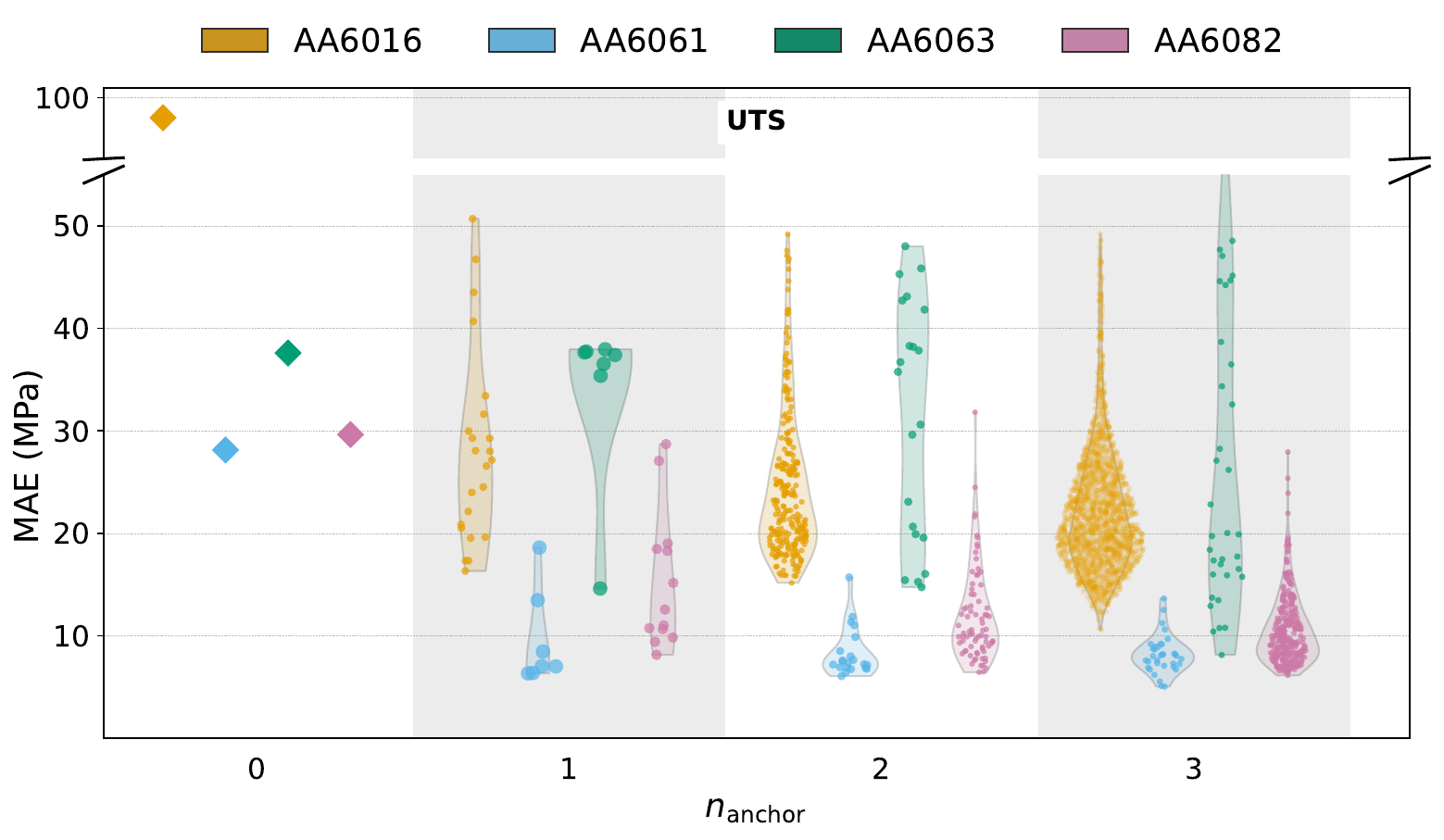}
    \caption{MAE-distribution for ultimate tensile strength as a function of the number of anchor conditions ($n_\mathrm{anchor}$) in leave-one-alloy-out cross validation. Scatter points show all possible combinations of anchor samples from the validation set. Note y-axis break.}
    \label{fig:loao_results_uts}
\end{figure}

Figure~\ref{fig:loao_results_uts} depicts the MAE-distribution for ultimate tensile strength as a function of the number of anchor conditions ($n_\mathrm{anchor}$) included in the training fold, displaying similar characteristics as for predicting yield strength (Figure~\ref{fig:loao_results}).

\subsection*{Parity Plots}
\begin{figure}[h!]
    \centering
    \includegraphics[width=\textwidth]{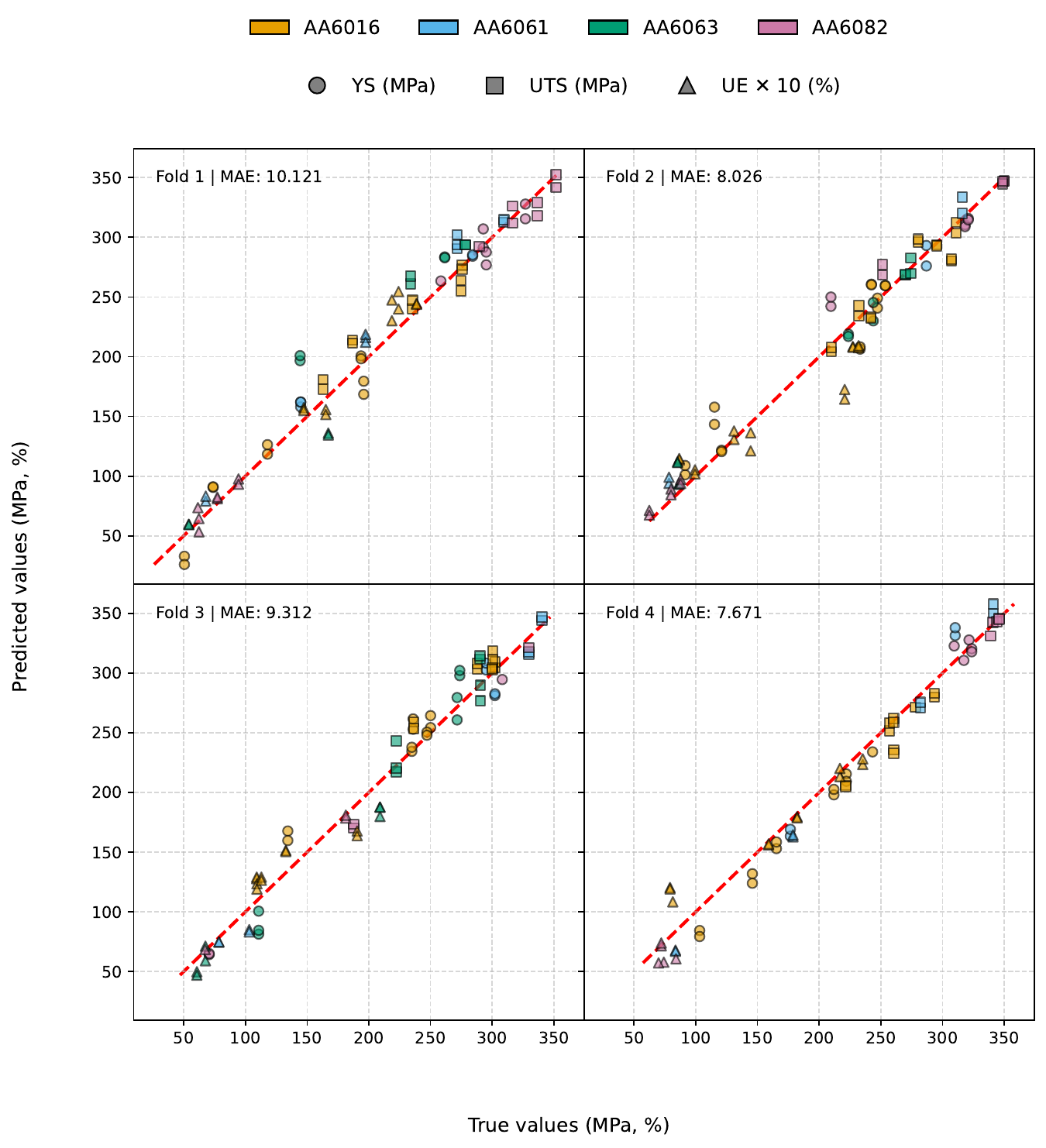}
    \caption{Parity plots for each fold in a 4-fold CV run, plotting predicted and true values of each condition and target property in the respective validation fold, with UE scaled by 10 for better readability.}
    \label{fig:parity}
\end{figure}

Figure~\ref{fig:parity} displays predicted and true values of each condition and target property in the held-out validation fold, for a single 4-fold CV run.

\clearpage~

\printbibliography

@article{abdi2010partial,
  title={Partial least squares regression and projection on latent structure regression ({PLS} Regression)},
  author={Abdi, Herv{\'e}},
  journal={Wiley interdisciplinary reviews: computational statistics},
  volume={2},
  number={1},
  pages={97--106},
  year={2010},
  publisher={Wiley Online Library}
}

@article{Arnoldt2024,
author = {Arnoldt, Aurel and Grohmann, Lukas and Strommer, Stephan and Österreicher, Johannes},
year = {2024},
month = {04},
pages = {},
title = {Differential scanning calorimetry of age-hardenable aluminium alloys: effects of sample preparation, experimental conditions, and baseline correction},
volume = {149},
journal = {Journal of Thermal Analysis and Calorimetry},
doi = {10.1007/s10973-024-13019-5}
}

@article{banhart2010natural,
  title={Natural aging in {Al-Mg-Si} alloys--a process of unexpected complexity},
  author={Banhart, John and Chang, Cynthia Sin Ting and Liang, Zeqin and Wanderka, Nelia and Lay, Matthew DH and Hill, Anita J},
  journal={Advanced engineering materials},
  volume={12},
  number={7},
  pages={559--571},
  year={2010},
  publisher={Wiley Online Library}
}

@article{edwards1998,
  title={The precipitation sequence in {Al--Mg--Si} alloys},
  author={Edwards, Geoffrey A and Stiller, Krystyna and Dunlop, Gordon L and Couper, Malcolm J},
  journal={Acta materialia},
  volume={46},
  number={11},
  pages={3893--3904},
  year={1998},
  publisher={Elsevier}
}

@article{Engler2019,
  title   = {Effect of natural ageing or pre-ageing on the evolution of precipitate structure and strength during age hardening of {Al–Mg–Si} alloy {AA} 6016},
  author  = {Engler, O. and Marioara, C. D. and Aruga, Y. and Kozuka, M. and Myhr, O. R.},
  journal = {Materials Science and Engineering: A},
  volume  = {759},
  pages   = {520--529},
  year    = {2019},
  doi     = {10.1016/j.msea.2019.05.066}
}

@article{falkinger2024modeling,
  title={Modeling the concurrent growth of inter-and intragranular Si precipitates during slow cooling of the alloy AA6016},
  author={Falkinger, Georg and Kahlenberg, Robert and Theissing, Moritz and Mitsche, Stefan and Thum, Angela and Pogatscher, Stefan},
  journal={European Journal of Materials},
  volume={4},
  number={1},
  pages={2316914},
  year={2024},
  publisher={Taylor \& Francis}
}

@article{falkinger2022analysis,
  title={Analysis of the evolution of Mg2Si precipitates during continuous cooling and subsequent re-heating of a 6061 aluminum alloy with differential scanning calorimetry and a simple model},
  author={Falkinger, Georg and Reisecker, Christian and Mitsche, Stefan},
  journal={International Journal of Materials Research},
  volume={113},
  number={4},
  pages={316--326},
  year={2022},
  publisher={de Gruyter}
}

@article{Grohmann2026,
  title={Age-hardening model for Al-Mg-Si alloys: a computationally efficient multi-phase approach to predict the evolution of yield strength},
  author={Grohmann, Lukas and Strommer, Stephan and Arnoldt, Aurel R and {\"O}sterreicher, Johannes A and Steinboeck, Andreas and Kugi, Andreas},
  journal={Mathematical and Computer Modelling of Dynamical Systems},
  volume={32},
  number={1},
  pages={2618478},
  year={2026},
  publisher={Taylor \& Francis},
  doi={10.1080/13873954.2026.2618478}
}

@article{jolliffe2016principal,
  title={Principal component analysis: a review and recent developments},
  author={Jolliffe, Ian T and Cadima, Jorge},
  journal={Philosophical transactions of the royal society A: Mathematical, Physical and Engineering Sciences},
  volume={374},
  number={2065},
  pages={20150202},
  year={2016},
  publisher={the Royal Society publishing}
}

@article{Kahlenberg2024,
  title   = {Revisiting high-energy X-ray diffraction and differential scanning calorimetry data of EN AW-6082 with mean field simulations},
  author  = {Kahlenberg, Robert and Schuster, Roman and Garc{\'i}a Arango, Nicol{\'a}s and Falkinger, Georg and Stark, Andreas and Milkereit, Benjamin and Kozeschnik, Ernst},
  journal = {Thermochimica Acta},
  volume  = {740},
  pages   = {179848},
  year    = {2024},
  doi     = {10.1016/j.tca.2024.179848}
}

@article{kahlenberg2025deconvolution,
  title={Deconvolution of superimposing reaction signals from DSC curves in selected Al-Mg-Si-(Cu) alloys by mean-field modeling and HEXRD},
  author={Kahlenberg, Robert and Falkinger, Georg and Schuster, Roman and Miesenberger, Bernhard and Arango, Nicol{\'a}s Garc{\'\i}a and Maawad, Emad and Povoden-Karadeniz, Erwin and Milkereit, Benjamin and Kozeschnik, Ernst},
  journal={Thermochimica Acta},
  pages={180181},
  year={2025},
  publisher={Elsevier}
}

@incollection{kaufman2013,
  author       = {J. Gilbert Kaufman},
  title        = {Understanding the Aluminum Temper Designation System},
  booktitle    = {Introduction to Aluminum Alloys and Tempers},
  publisher    = {ASM International},
  year         = {2013},
  note         = {Chapter available online at NIST Materials Data Repository},
  url          = {https://materialsdata.nist.gov/handle/11115/186},
}

@article{lai2017low,
  title={Low-alloy-correlated reversal of the precipitation sequence in Al-Mg-Si alloys},
  author={Lai, YX and Jiang, BC and Liu, CH and Chen, ZK and Wu, CL and Chen, JH},
  journal={Journal of Alloys and Compounds},
  volume={701},
  pages={94--98},
  year={2017},
  publisher={Elsevier}
}

@article{Li2022,
  title   = {Synergistic improvement in bake-hardening response and natural aging stability of Al–Mg–Si–Cu–Zn alloys via non-isothermal pre-aging treatment},
  author  = {Li, Gaojie and Guo, Mingxing and Du, Jinqing and Zhuang, Linzhong},
  journal = {Materials \& Design},
  volume  = {218},
  pages   = {110714},
  year    = {2022},
  doi     = {10.1016/j.matdes.2022.110714}
}

@article{madanat2025heat,
  title={The Heat Treatment Route Impact on 6063 Al-Mg-Si Alloy: Implications for Mechanical Properties},
  author={Madanat, Madanat and Al-Masri, Qutaibah and Arabiat, Ayeh and Alrwashdeh, Saad and Mousa, Marwan},
  journal={Jordan Journal of Physics},
  volume={18},
  number={3},
  pages={319--328},
  year={2025}
}

@article{marioara2024atomic,
  title={Atomic structure of clusters and GP-zones in an Al-Mg-Si alloy},
  author={Marioara, Calin D and Andersen, Sigmund J and Hell, Christoph and Frafjord, Jonas and Friis, Jesper and Bj{\o}rge, Ruben and Ringdalen, Inga G and Engler, Olaf and Holmestad, Randi},
  journal={Acta Materialia},
  volume={269},
  pages={119811},
  year={2024},
  publisher={Elsevier}
}

@article{milkereit2015quench,
  title={Quench sensitivity of Al--Mg--Si alloys: A model for linear cooling and strengthening},
  author={Milkereit, B and Starink, MJ},
  journal={Materials \& Design},
  volume={76},
  pages={117--129},
  year={2015},
  publisher={Elsevier}
}

@article{Murakami2025,
  title   = {Mechanical property prediction of aluminium alloys with varied silicon content using deep learning},
  author  = {Murakami, Yuichiro and Furushima, Ryoichi and Shiga, Keiji and Miyajima, Tatsuya and Omura, Naoki},
  journal = {Acta Materialia},
  volume  = {286},
  pages   = {120683},
  year    = {2025},
  doi     = {10.1016/j.actamat.2024.120683}
}

@article{pan2009survey,
  title={A survey on transfer learning},
  author={Pan, Sinno Jialin and Yang, Qiang},
  journal={IEEE Transactions on knowledge and data engineering},
  volume={22},
  number={10},
  pages={1345--1359},
  year={2009},
  publisher={IEEE}
}

@article{Park2023,
  title   = {Full-range stress--strain curve estimation of aluminum alloys using machine learning-aided ultrasound},
  author  = {Park, Seong-Hyun and Chung, Junyeon and Yi, Kiyoon and Sohn, Hoon and Jhang, Kyung-Young},
  journal = {Ultrasonics},
  volume  = {135},
  pages   = {107146},
  year    = {2023},
  doi     = {10.1016/j.ultras.2023.107146}
}

@article{park2025three,
  author  = {Park, Seong-Hyun},
  title   = {Three-dimensional stress-strain curve estimation and visualization using ultrasound and the Ramberg-Osgood model: A nondestructive approach to material characterization},
  journal = {Mechanical Systems and Signal Processing},
  volume  = {224},
  pages   = {112087},
  year    = {2025}
}

@software{pichlmann_2026_18936092,
author       = {Pichlmann, Lukas and
              Studer, Samuel},
title        = {DSCorr: Correlating DSC measurements and
               mechanical properties in 6xxx aluminum alloys
              },
month        = mar,
year         = 2026,
publisher    = {Zenodo},
version      = {v1.0.0},
doi          = {10.5281/zenodo.18936092},
url          = {https://doi.org/10.5281/zenodo.18936092},
swhid        = {swh:1:dir:1875ee3cdd8caafa352307cd1c60d7193d2b2813
               ;origin=https://doi.org/10.5281/zenodo.18936091;vi
               sit=swh:1:snp:1b1204ec2fa955afb83e789a3c6b16fa997d
               2613;anchor=swh:1:rel:9375dea8cecc10761a4f9cb75271
               e160b0f28822;path=dscorr
              },
}

@article{poole2005,
  title={The shearable--non-shearable transition in Al--Mg--Si--Cu precipitation hardening alloys: implications on the distribution of slip, work hardening and fracture},
  author={Poole*, WJ and Wang, X and Lloyd, DJ and Embury, JD},
  journal={Philosophical Magazine},
  volume={85},
  number={26-27},
  pages={3113--3135},
  year={2005},
  publisher={Taylor \& Francis}
}

@article{robson2020advances,
  title={Advances in microstructural understanding of wrought aluminum alloys},
  author={Robson, JD and Engler, O and Sigli, C and Deschamps, Alexis and Poole, WJ},
  journal={Metallurgical and Materials Transactions A},
  volume={51},
  number={9},
  pages={4377--4389},
  year={2020},
  publisher={Springer}
}

@article{scikit-learn,
  title={Scikit-learn: Machine Learning in {P}ython},
  author={Pedregosa, F. and Varoquaux, G. and Gramfort, A. and Michel, V.
          and Thirion, B. and Grisel, O. and Blondel, M. and Prettenhofer, P.
          and Weiss, R. and Dubourg, V. and Vanderplas, J. and Passos, A. and
          Cournapeau, D. and Brucher, M. and Perrot, M. and Duchesnay, E.},
  journal={Journal of Machine Learning Research},
  volume={12},
  pages={2825--2830},
  year={2011}
}

@article{tibshirani1996regression,
  title={Regression shrinkage and selection via the lasso},
  author={Tibshirani, Robert},
  journal={Journal of the Royal Statistical Society Series B: Statistical Methodology},
  volume={58},
  number={1},
  pages={267--288},
  year={1996},
  publisher={Oxford University Press}
}

@article{vecchio2021,
  title={High-throughput rapid experimental alloy development (HT-READ)},
  author={Vecchio, Kenneth S and Dippo, Olivia F and Kaufmann, Kevin R and Liu, Xiao},
  journal={Acta Materialia},
  volume={221},
  pages={117352},
  year={2021},
  publisher={Elsevier}
}

@article{wold2001pls,
  title={PLS-regression: a basic tool of chemometrics},
  author={Wold, Svante and Sj{\"o}str{\"o}m, Michael and Eriksson, Lennart},
  journal={Chemometrics and intelligent laboratory systems},
  volume={58},
  number={2},
  pages={109--130},
  year={2001},
  publisher={Elsevier}
}

@article{xu2023small,
  title={Small data machine learning in materials science},
  author={Xu, Pengcheng and Ji, Xiaobo and Li, Minjie and Lu, Wencong},
  journal={npj Computational Materials},
  volume={9},
  number={1},
  pages={42},
  year={2023},
  publisher={Nature Publishing Group UK London}
}

@article{Yang2021,
  title   = {Natural and artificial ageing in aluminium alloys – the role of excess vacancies},
  author  = {Yang, Zi and Banhart, John},
  journal = {Acta Materialia},
  volume  = {215},
  pages   = {117014},
  year    = {2021},
  doi     = {10.1016/j.actamat.2021.117014}
}

\end{document}